\documentclass[aps,showpacs,prd,twocolumn,nofootinbib,nobibnotes]{revtex4-2}

\usepackage{color}
\usepackage{amsmath}
\usepackage{amssymb}
\usepackage[utf8]{inputenc}
\usepackage{amsfonts}
\usepackage{bm}
\usepackage{bbold}
\usepackage{graphics}
\usepackage{graphicx}

\newcommand{\be}{\begin{equation}}
\newcommand{\ee}{\end{equation}}
\newcommand{\ba}{\begin{eqnarray}}
\newcommand{\ea}{\end{eqnarray}}

\def\ni{\noindent}

\begin{document}

\title{\Large Dispersion and absorption effects in the linearized Euler-Heisenberg electrodynamics under
an external magnetic field}

\author{G. R. Santos} \email{guirafael.ufrrj@gmail.com}
\affiliation{Departamento de F\'{i}sica, Universidade Federal Rural do Rio de Janeiro, BR 465-07, 23890-971, Serop\'edica, RJ, Brazil}

\author{M. J. Neves} \email{mariojr@ufrrj.br}
%\affiliation{Department of Physics and Astronomy, University of Alabama, Tuscaloosa, Alabama 35487, USA}
\affiliation{Departamento de F\'{i}sica, Universidade Federal Rural do Rio de Janeiro, BR 465-07, 23890-971, Serop\'edica, RJ, Brazil}

%\author{Patricio Gaete} \email{patricio.gaete@usm.cl}
%\affiliation{Departamento de F\'{i}sica and Centro Cient\'{i}fico-Tecnol\'ogico de Valpara\'{i}so-CCTVal,
%Universidad T\'{e}cnica Federico Santa Mar\'{i}a, Valpara\'{i}so, Chile}

%\author{ L. P. R. Ospedal }  \email{ leoopr@cbpf.br}
%\affiliation{Centro Brasileiro de Pesquisas F\'isicas, Rua Dr. Xavier Sigaud
%150, Urca, Rio de Janeiro, Brasil, CEP 22290-180}

%
\date{\today}

\begin{abstract}
\ni

The effects of the Ohmic and magnetic density currents are investigated in the linearized Euler-Heisenberg electrodynamics. The linearization is introduced through an external magnetic field, in which the vector potential of the Euler-Heisenberg electrodynamics is expanded around of a magnetic background field, that we consider uniform and constant in this paper. From the Euler-Heisenberg linearized equations, we obtain the solutions for the refractive index associated with the electromagnetic wave superposition, when the current density is ruled by the Ohm law, and in the second case, when the current density is set by a isotropic magnetic conductivity. These solutions are functions of the magnetic background $({\bf B})$, of the wave propagation direction $({\bf k})$, it also depends on the conductivity, and on the wave frequency. As consequence, the dispersion and the absorption of plane waves change when ${\bf B}$ is parallel to ${\bf k}$ in relation to the case of ${\bf B}$ perpendicular to ${\bf k}$ in the medium. The characteristics of the refraction index related to directions of ${\bf B}$ and of the wave polarization open a discussion for the birefringence in this medium.

\end{abstract}

%\maketitle

\pagestyle{myheadings}

\maketitle

\pagestyle{myheadings}
\markright{Dispersion and absorption effects in the linearized EH ED under an external magnetic field}

\section{Introduction}
The Euler-Heisenberg (EH) electrodynamics (ED) was the first non-linear ED discovered through the radiative corrections of the quantum
electrodynamics, when it is submitted to an external electromagnetic field \cite{EH}. For a historical review in EH ED,
see \cite{Dunne}. As a second example, some earlier years, the Born-Infeld (BI) electrodynamics (ED) is one of most famous non-linear 
extensions of the Maxwell ED in the literature \cite{Born,BornNature}. Originally, it was proposed to explain the classical electron self-energy,
since that, in Maxwell ED, the electric field for a rest like-point charge is not defined at the origin. Nowadays, several other examples
of non-linear EDs have applications or solutions in many research areas, as anomalous couplings in physics beyond the Standard Model,
black holes, string theory, Dirac materials and others, see the refs. \cite{Ellis,Sorokin,Tseytlin,Fradkin,Pope,Keser,Zhao,Kiril,Kruglov,Garcia}.
The investigation of phenomena under action of an external field also is a subject of interest in non-linear EDs \cite{Birula,Liao}.
An interesting application of non-linear EDs is the case of astrophysical objects, as neutron stars, in which the magnetic field has strong magnitude 
in the range of $\sim 10^{4} - 10^{11}$ T \cite{Andreas1,Andreas2}. Thereby, the introduction of the external electromagnetic field in non-linear EDs 
is an approach that allows the investigation of propagation effects, as the dispersion relations, group velocities, the refractive index, and also the
characteristics of the material medium under an external (uniform and constant) magnetic field \cite{MJNevesEDN,PaixaoJHEP,MJNevesPRD2023}.
The birefringence phenomenon also emerge in some non-linear EDs, see \cite{Kruglov2007,Kruglov2015,Kruglov2010}.
For a complete description of the polarized vacuum with laser (PVLAS) experiment to measure the vacuum birefringence, see \cite{25years}. 
The study of electromagnetic waves in materials, as conductors, is one of known applications of the Ohm law at room temperature
\cite{Zangwill}. It allows to obtain the dispersion and absorption of waves in the material medium. Other current density
discussed in the literature of material physics is known as the magnetic current density. The current density vector is proportional 
to the magnetic field, in which the proportionality constant is called magnetic conductivity \cite{Kharzeev1,Kharzeev2,Kharzeev3}. This current density
has origin from the systems with asymmetry of left- and right-handed chiral fermions, that is known as the Chiral Magnetic Effect.
In Weyl semimetals, the CME is related to massless fermions acquire velocity along the magnetic field \cite{Burkov,Pedro}.
Meeting all these motivations, we investigate the dispersion and absorption of the wave propagation in the linearized EH ED
by an external magnetic field, when the material medium is governed by the Ohm law current density, and posteriorly,
when the current density is proportional to the magnetic field. The refractive index of the material medium depends on the conductivity,
on the wave frequency, and also on the magnetic background. Our main motivation is the investigation of the birefringence phenomenon
through the Own law and the non-linearity type EH ED. Since we consider the EH ED as a non-linear classical ED throughout the manuscript. We show as the birefringence for wave plane emerges as consequence of the conductivity, and of the external magnetic field. The paper is organized as follows : In the section \ref{sec2}, we show the linearization of the EH ED in the presence of an external (uniform and constant) magnetic field. In the section \ref{sec3}, we obtain the dispersion and absorption of waves for the case of a Ohmic current density. The section \ref{sec4} is dedicated to the wave dispersion for an isotropic magnetic current density. In the section \ref{sec5}, we show the birefringence phenomenon associated with the solutions of the refractive index in the Ohm law. For end, the final considerations are cast in the section \ref{sec6}.

Throughout this work, we adopt natural units $\hbar=c=1$, with $\epsilon_0 = \mu_0 = 1$. We use the conversion $1\,\mbox{m}=5 \times 10^{12} \, \mbox{MeV}^{-1}$ for a physical quantity with length dimension. The electric and magnetic fields have squared-energy mass dimension, where the conversion of Volt/m and Tesla (T) to the natural units is given by $1 \, \mbox{Volt/m}=2.27 \times 10^{-18} \, \mbox{MeV}^2$ and $1 \, \mbox{T} =  6.8 \times 10^{-10} \, \mbox{MeV}^2$, respectively. The signature of the metric in the paper is $\eta_{\mu\nu}=\mbox{diag}(+1,-1,-1,-1)$.

\section{The linearized Euler-Heisenberg electrodynamics}
\label{sec2}

The non-linear EH ED in the presence of a source $J^{\mu}=(\rho,{\bf J})$ is governed by the lagrangian density
\begin{eqnarray}\label{Lmodel}
{\cal L}_{EH}= {\cal F}_{0}+\frac{2\alpha^2}{45m^4} \left(\, 4 \, {\cal F}_{0}^2\,+\,7 \, {\cal G}_{0}^2 \,\right)-J_{\mu}A_{0}^{\;\mu} \; ,
\end{eqnarray}
where $J^{\mu}=(\rho,{\bf J})$ is a $4$-current density, ${\cal F}_0$ and ${\cal G}_0$ are the gauge and Lorentz invariant
\begin{subequations}
\begin{eqnarray}
{\cal F}_{0}&=&-\frac{1}{4} \, F_{0\mu\nu}^{2}=\frac{1}{2} \, \left( {\bf E}_{0}^2-{\bf B}_{0}^2\right) \; ,
\\
{\cal G}_{0}&=&-\frac{1}{4} \, F_{0\mu\nu}\widetilde{F}_{0}^{\;\,\,\mu\nu}={\bf E}_{0}\cdot{\bf B}_{0} \; ,
\end{eqnarray}
\end{subequations}
and $\alpha=e^2=(137)^{-1}=0.00729$ is the fine structure constant, and $m=0.5 \, \mbox{MeV}$ is the electron mass.
The non-linear effects of a classical electrodynamics described by the lagrangian density are sensible when the EM field
is in the range of the Schwinger's critical field $B_{c}=m^2/e=0.82 \, \mbox{MeV}^2=1.22 \times 10^{9}$ T.

%$\beta$ is the BI parameter with electromagnetic field dimension (energy to the squared in natural units). The $\beta$-BI parameter is interpreted as a critical field of the theory in which it is reduced to the usual Maxwell ED in the limit $\beta \rightarrow \infty$.
%For the analysis of propagation effects, we use $\beta=1.187 \times 10^{20} \, \mbox{V}/\mbox{m}\simeq 256 \; \mbox{MeV}^2$ associated
%with the finite self-energy for the electron \cite{Born}.

%
The prescription for the external and uniform magnetic field is introduced through the gauge $4$-potential $A_{0\mu}=a_{\mu}+A_{B\mu}$, where $a_{\mu}$ is the propagating $4$-potential in the space-time, and $A_{B\mu}$ is the potential associated with the magnetic background field ${\bf B}$. The field-strength tensor is also decomposed as $F_{0\mu\nu}=f_{\mu\nu}+F_{B\mu\nu}$, in which $f^{\mu\nu}=\partial^{\mu}a^{\nu}-\partial^{\nu}a^{\mu}=\left( \, -e^{i} \, , \, -\epsilon^{ijk} \, b^{k} \, \right)$
denotes the EM field strength tensor of the propagating EM fields, whereas $F_{B}^{\;\,\mu\nu}=\partial^{\mu}A_{B}^{\;\,\nu}-\partial^{\nu}A_{B}^{\;\,\mu} =\left( \, 0 \, , \, -\epsilon^{ijk} \, B^{k} \, \right)$ sets the field strength of the magnetic background. Using this approach, the lagrangian density (\ref{Lmodel}) up to second order in the propagation gauge field is read below
\begin{eqnarray}\label{Lmodel2}
{\cal L}_{EH}^{(2)} &=& -\frac{1}{4} \, c_{1} \, f_{\mu\nu}^{\, 2}
-\frac{1}{4} \, c_{2} \, f_{\mu\nu} \, \widetilde{f}^{\mu\nu}
\nonumber \\
&&
\hspace{-0.5cm}
+\,\frac{1}{8} \, Q_{B\mu\nu\kappa\lambda} \, f^{\mu\nu} \, f^{\kappa\lambda}
-J_{\mu}\,(a^{\mu}+A_{B}^{\mu}) \; ,
\end{eqnarray}
where $\widetilde{f}_{\mu\nu}=\varepsilon_{\mu\nu\alpha\beta}f^{\alpha\beta}/2$ is the strength field dual tensor, and we have defined the tensor $Q_{B\mu\nu\kappa\lambda}$ evaluated at the magnetic background as follows
\begin{eqnarray}
Q_{B\mu\nu\kappa\lambda} \!&=&\! d_{1} \, F_{B\mu\nu}F_{B\kappa\lambda}
+d_{2} \, \widetilde{F}_{B\mu\nu}\widetilde{F}_{B\kappa\lambda}
\nonumber \\
&&
\hspace{-0.3cm}
+\,d_{3} \, F_{B\mu\nu}\widetilde{F}_{B\kappa\lambda}
+ d_{3} \, \widetilde{F}_{B\mu\nu} F_{B\kappa\lambda} \; .
\end{eqnarray}
%
%
%with $\widetilde{F}_{B\mu\nu}=\varepsilon_{\mu\nu\alpha\beta}\,F_{B}^{\;\,\,\alpha\beta}/2$ being} the dual tensor of the background field.
The coefficients of the expansion $c_{i} \, (i=1,2)$ and $d_{i} \, (i=1,2,3)$ are defined by
\begin{eqnarray}\label{coefficientsMN}
c_{1}=\left.\frac{\partial{\cal L}}{\partial{\cal F}_0}\right|_{{\bf B}} ,
\left. c_{2}=\frac{\partial{\cal L}}{\partial{\cal G}_0}\right|_{{\bf B}} ,
%\nonumber \\
\left. d_{1}=\frac{\partial^2{\cal L}}{\partial{\cal F}_0^2}\right|_{{\bf B}} ,
\nonumber \\
\left. d_{2}=\frac{\partial^2{\cal L}}{\partial{\cal G}_0^2}\right|_{{\bf B}} ,
%\nonumber \\
\left. d_{3}=\frac{\partial^2{\cal L}}{\partial{\cal F}_0\partial{\cal G}_0}\right|_{{\bf B}} \; .
\end{eqnarray}
%
%

%We now examine wave propagation effects under a uniform magnetic field for the
%non-linear ED (\ref{Lmodel}). Follo\-wing a similar procedure of ref. \cite{MJNevesEDN},

%
Substituting the lagrangian (\ref{Lmodel}), the EH
coefficients in a uniform magnetic background are given by
\begin{subequations}
\begin{eqnarray}\label{coeficientsresult}
c_{1} &=& 1-\frac{8\alpha^2 \, {\bf B}^2}{45m^4}
\hspace{0.3cm} , \hspace{0.3cm}
c_{2} = 0 \; ,
\\
d_{1} &=& \frac{16\alpha^2}{45m^{4}}
\; ,
\\
d_{2} &=& \frac{28\alpha^2}{45m^{4}}
\hspace{0.3cm} , \hspace{0.3cm}
d_{3}=0 \; .
\end{eqnarray}
\end{subequations}
The non-null coefficients simplifies the lagrangian density (\ref{Lmodel2}) as
\begin{eqnarray}\label{L2}
{\cal L}_{EH}^{(2)} &=& -\frac{1}{4} \, c_{1} \, f_{\mu\nu}^{\, 2}
+\frac{d_1}{8} \, (F_{B\mu\nu}f^{\mu\nu})^{2}
+\frac{d_2}{8} \, (\widetilde{F}_{B\mu\nu}f^{\mu\nu})^{2}
\nonumber \\
&&
-J_{\mu}\,(a^{\mu}+A_{B}^{\mu}) \; .
\end{eqnarray}
In the limit $\alpha \rightarrow 0$, the coefficients are reduced to $c_1=1$ and $d_1=d_2=0$, and the lagrangian (\ref{L2}) leads to the Maxwell ED for the propagating fields. Since that the magnetic background field is constant and uniform, in this particular case the coefficients of the expansion do not depend on the space-time coordinates.
The action principle applied to the lagrangian (\ref{L2}), in relation to $a^{\mu}$, yields the linearized field equation
\begin{equation}\label{eqf}
\hspace{-0.5cm}
\partial^{\mu}G_{\mu\nu}=J_{\nu} \; ,
\end{equation}
where $G_{\mu\nu}=(c_{1}\,\eta_{\kappa\mu}\eta_{\lambda\nu}
- d_1 \, F_{B\mu\nu} F_{B\kappa\lambda}/2
-d_2 \, \widetilde{F}_{B\mu\nu}\widetilde{F}_{B\kappa\lambda}/2)f^{\kappa\lambda}$,
and the dual tensor $\widetilde{f}^{\mu\nu}$ satisfies the Bianchi identity $\partial_{\mu}\widetilde{f}^{\mu\nu}=0$. The quadri-current satisfies the charge conservation equation $\partial_{\mu}J^{\mu}=0$.
In vector notation, the correspondent field equations in the presence of a charge density $\rho$, and of a current density ${\bf J}$, 
are
\begin{subequations} \label{eq.linea}
\begin{eqnarray}
\nabla\cdot{\mathbf{e}} + f \; \mathbf{B} \cdot \nabla(\mathbf{B} \cdot \mathbf{e}) \!= \rho \, ,
\label{eqdive}
\\
\nabla\times{\mathbf{e}}+\frac{\partial{\mathbf{b}}}{\partial t} = {\mathbf{0}} \;
\label{eqd rot e}
\hspace{0.2cm} , \hspace{0.2cm}
\nabla\cdot{\mathbf{b}} = 0 \,
\hspace{0.2cm} , \hspace{0.2cm}
\\
\nabla\times{\mathbf{b}} + d \; {\mathbf{B}} \times \nabla(\mathbf{B} \cdot {\mathbf{b}}) = {\mathbf{J}} + \frac{\partial{\mathbf{e}}}{\partial t}+
\nonumber \\
+  f \, \mathbf{B} \, \frac{\partial}{\partial t}({\mathbf{B}}\cdot{\mathbf{e}}) \; ,
\label{eqrotb}
\end{eqnarray}
\end{subequations}
in which we rewrite the coefficients as
\begin{subequations}
\begin{eqnarray}
d \!&=&\! \frac{d_1}{c_1}\simeq \frac{16\alpha^2}{45m^{4}}= 3 \times 10^{-4}\, \mbox{MeV}^{-4} \; ,
\\
%\hspace{0.3cm} \mbox{and} \hspace{0.3cm}
f \!&=&\! \frac{d_2}{c_1}\simeq\frac{28\alpha^2}{45m^{4}} = 5.29 \times 10^{-4}\, \mbox{MeV}^{-4} \; ,
\end{eqnarray}
\end{subequations}
that implies into the relation $d\simeq4f/7$. The limit $\alpha \rightarrow 0$ recovers the usual Maxwell equations in (\ref{eqdive})-(\ref{eqrotb}) for
${\bf e}$ and ${\bf b}$. The ${\bf B}$-magnetic field is so interpreted as a background vector in all the previous
equations. The presence of this background field modifies the dispersion relations associated with the plane wave solutions.
It will be explored in the next section for a Ohmic current density.

\section{The dispersion and absorption in the presence of Ohmic current}
\label{sec3}
Since it is known in classical electrodynamics for a class of materials, such as the conductors, the current density is governed by the
Ohm law
\begin{eqnarray}\label{Johm}
\mathbf{J} = \sigma \, \mathbf{e} \; ,
\end{eqnarray}
where $\sigma$ is the electric conductivity at room temperature, that is characteristics of the material medium.
For the analysis of the plane waves, we substitute the Fourier transforms in the linearized equations (\ref{eqdive})-(\ref{eqrotb})
\begin{subequations}	
\begin{eqnarray}\label{Ondas Planas}
{\mathbf{e}}(x) = \int \frac{d^4k}{(2\pi)^4} \; {\mathbf{e}}_0(k) \; e^{-ik\cdot x} \; ,
\\
{\mathbf{b}}(x) = \int \frac{d^4k}{(2\pi)^4} \; {\mathbf{b}}_0(k) \; e^{-ik\cdot x} \; ,
\\
\rho(x)=\int \frac{d^4k}{(2\pi)^4} \; \rho_{0}(k) \; e^{-ik\cdot x} \; ,
\end{eqnarray}
\end{subequations}
where the scalar product is $k\cdot x=\omega\,t-{\bf k}\cdot{\bf r}$, in which ${\bf k}$ is the wave vector,
$\omega$ is the wave frequency, ${\bf e}_0(k)$ and ${\bf b}_0(k)$ are the electric and magnetic wave amplitudes,
respectively, and $\rho_{0}(k)$ is Fourier transform of the charge density.
Thereby, we obtain the equations fields in the momentum space
\begin{subequations}
\begin{eqnarray}
\mathbf{k} \cdot {\mathbf{e_0}} + f \; (\mathbf{B} \cdot \mathbf{k}) (\mathbf{B} \cdot \mathbf{e_0}) \!= \rho_{0}(k) \, ,
\label{eqke0}
\hspace{0.5cm}
\\
\hspace{-2cm} \mathbf{k} \times {\mathbf{e_0}}  = \omega \, \mathbf{b_0} \;
\label{eqkb0}
\hspace{0.2cm} , \hspace{0.2cm}
\mathbf{k} \cdot {\mathbf{b_0}} = 0 \; ,
\hspace{1.2cm}
\\
\mathbf{k} \times {\mathbf{b_0}} +  \frac{4f}{7}\,(\mathbf{B} \times \mathbf{k}) \, (\mathbf{B} \cdot {\mathbf{b_0}}) = - i \,\sigma \, {\mathbf{e_0}} \nonumber
\\
- f \, \omega \, \mathbf{B} \, (\mathbf{B} \cdot \mathbf{e_0}) -\omega \, \mathbf{e_0} \; .
\hspace{1.0cm}
\label{eqkxb0}
\end{eqnarray}
\end{subequations}
The equations (\ref{eqkb0})-(\ref{eqkxb0}) can be combined such that we obtain the wave equation for
the electric amplitude $e_{0i} \, (i=1,2,3)$ :
\begin{eqnarray}\label{EqMij}
M_{ij} \, e_{0j} = 0 \; ,
\end{eqnarray}
where $M_{ij}$ is the wave matrix
%
%
%Nesse momento, será transferido para a forma de componentes,
%	
%\begin{eqnarray}
%\mathbf{k}(\mathbf{k} \cdot \mathbf{e}_0) + \mathbf{e}_{0i} (\omega^2 - \,\mathbf{k}^2 +  i \, \sigma \, \omega) + ({\mathbf{B}}^2+\beta^2)^{-1} \, (\mathbf{B} \times \mathbf{k})_i \, [\mathbf{e}_0 \cdot (\mathbf{B} \times \mathbf{k})]+  \beta^2 \, \omega^2 \, \mathbf{B}_i \, (\mathbf{B} \cdot \mathbf{e_0}) \!\!\!&=&\!\!\! \, 0 , \hspace{1.2 cm}
%\end{eqnarray}
%	
%por fim, coloca-se o $\mathbf{e}_{0j}$ em evidência,
%	
%	 \begin{eqnarray}
%	  [\mathbf{k}_i \, \mathbf{k}_j + \mathbf{e}_{0i} (\omega^2 - \,\mathbf{k}^2 +  i \, \sigma \, \omega) +  ({\mathbf{B}}^2+\beta^2)^{-1} \, (\mathbf{B} \times \mathbf{k})_i \, (\mathbf{B} \times \mathbf{k})_j +   \beta^2 \, \omega^2 \, \mathbf{B}_i \, \mathbf{B}_j]\, \mathbf{e}_{0j} \!\!\!\!&=&\!\!\!\! 0 \,. \hspace{1.2 cm}
%\end{eqnarray}	
%	
%	Dividindo toda a equação por $\omega^2$, para adquirir os índices de refração ($\mathbf{n}$) ,
%	
%	\begin{eqnarray}
%	[\mathbf{n}_i \, \mathbf{n}_j + \delta_{i\, j} (1 - \,\mathbf{n}^2 +  i \, \frac{\sigma}{\omega}) +  ({\mathbf{B}}^2+\beta^2)^{-1} \, (\mathbf{B} \times \mathbf{n})_i \, (\mathbf{B} \times \mathbf{n})_j +   \beta^2 \, \, \mathbf{B}_i \, \mathbf{B}_j]\,\mathbf{e}_{0j} \!\!\!&=&\!\!\! 0 \,. \hspace{0.5 cm}
%	\end{eqnarray}
%	
%Assim, define-se a equação $M_{i\,j} \, \mathbf{e}_{0j} = 0$, onde a matriz $M_{i \, j}$,
%	
\begin{eqnarray}\label{Mij}
M_{ij} = \delta_{i j} \left(1 - \mathbf{n}^2 +  i \, \frac{\sigma}{\omega}\right) + n_i \, n_j
\nonumber \\
+ \frac{4f}{7}\,(\mathbf{B} \times \mathbf{n})_i \, (\mathbf{B} \times \mathbf{n})_j + f  \, B_i \, B_j \; .
\end{eqnarray}
%
%escrita em termos das componentes $n_{i}=k_{i}/\omega$, na qual o índice de refração do meio é definido por $n=\sqrt{n_i\,n_i}$.
%As soluções não triviais de (\ref{Mij}) impõem que $\det(M_{ij})=0$, que leva às equações
%	
We have written the symmetric matrix $M_{ij}$ in terms of the refractive index components $n_{i}=k_{i}/\omega$, whose the refractive index is defined by $n=\sqrt{n_i\,n_i}$. 
The determinant of $M_{ij}$ is given by
%Não é interessante que o $\mathbf{e}_{0j}$ seja nulo, então o determinante da matriz $M_{i \,j}$, terá que ser nulo. Com isso, terá duas soluções diferentes para o índice de refração,
%	
% Então,
%	
%\begin{eqnarray}
	%a(a^2 + a(c \mathbf{B}^2 + b \mathbf{n}^2 + 2 d (\mathbf{B}^2 \cdot \mathbf{n}^2)+ b c - d^2 (\mathbf{B} \cdot \mathbf{n})^2 (\mathbf{B} \times \mathbf{n})^2) &=& 0\,,
	%\end{eqnarray}
%	
%onde, $a = 1 - \mathbf{n}^2 + ({\mathbf{B}}^2+\beta^2)^{-1} \, (\mathbf{B}^2 \times \mathbf{n}^2) + i \frac{\sigma}{\omega}$,  $\; b = ({\mathbf{B}}^2+\beta^2)^{-1} \, \mathbf{B}^2$ e $ \, c = \beta^2 - ({\mathbf{B}}^2+\beta^2)^{-1} \, \mathbf{n}^2$. Observe que têm duas soluções, uma para $a=0$ e a outra quando a expressão dentro dos parênteses for igual a zero, separando em duas soluções,
%	
%
%
\begin{eqnarray}\label{detMij}
\det(M_{ij})=\left[n^2-1-\frac{i\sigma}{\omega}-\frac{4f}{7}({\bf B}\times{\bf n})^2\right]
\nonumber \\
\times \left[\left(n^2-1-\frac{i\sigma}{\omega}\right)\left(1+\frac{i\sigma}{\omega}+f\,{\bf B}^2 \right)-f({\bf B}\times{\bf n})^2 \right] \; .
\hspace{-1.0cm}
\nonumber \\
\end{eqnarray}
The non-trivial solutions of (\ref{EqMij}) impose that $\det(M_{ij})=0$, where (\ref{detMij}) leads to the equations
\begin{subequations}
\begin{eqnarray}
&&
n_1^2-1-\frac{i\sigma}{\omega} = \frac{4f}{7}\, (\mathbf{B} \times \mathbf{n}_1)^2 \; ,
\label{n1}
\\
&&
\left(\mathbf{n}_2^2 - 1 - \frac{i\sigma}{\omega}\right) \left(1 + \frac{i\sigma}{\omega}+f\,{\bf B}^2 \right)
= f \left( \mathbf{B} \times \mathbf{n}_2\right)^2  \, . \label{n2} \nonumber \\
\end{eqnarray}
\end{subequations}
%	
%A solução de (\ref{n1}) é dada por
%	
The solution of (\ref{n1}) is given by
\begin{eqnarray}
n_1= \sqrt{\dfrac{1 + i\, \frac{\sigma}{\omega}}{1 - (4f/7) \, (\mathbf{B}\times\mathbf{\hat{k}})^2}} \; ,
\end{eqnarray}
that is not defined at $|{\bf B}\times\hat{{\bf k}}| = 57.51 \, \mbox{MeV}^{2}$.
If $|{\bf B}\times\hat{{\bf k}}| < 57.51 \, \mbox{MeV}^{2}$, the solution $n_1$ can be written as
\begin{equation}\label{n1ReIm}
n_1 = \Re[n_1]+i\,\Im[n_1] \; ,
\end{equation}
where the real and imaginary parts are given by
\begin{subequations}
\begin{eqnarray}
\Re[n_1] &=& \frac{1}{\sqrt{2}}\sqrt{ \dfrac{\sqrt{1+ \frac{\sigma^2}{\omega^2}}+1 }{ 1 - (4f/7) \, (\mathbf{B}\times\mathbf{\hat{k}})^2} } \; ,
\label{Ren1B}
\\
\Im[n_1] &=& \frac{1}{\sqrt{2}}\sqrt{ \dfrac{ \sqrt{1+ \frac{\sigma^2}{\omega^2}}-1 }{ 1 - (4f/7) \, (\mathbf{B}\times\mathbf{\hat{k}})^2 } }  \; .
\label{Imn1B}
\end{eqnarray}
\end{subequations}
On the other hand, if $|{\bf B}\times\hat{{\bf k}}| > 57.51 \, \mbox{MeV}^{2}$, the solution of (\ref{n1}) is $n_1^{\prime}=\Re[n_1^{\prime}]+i\,\Im[n_1^{\prime}]$, where
\begin{subequations}
\begin{eqnarray}
\Re[n_1^{\prime}] &=& -\frac{1}{\sqrt{2}} \sqrt{ \dfrac{\sqrt{1+ \frac{\sigma^2}{\omega^2}}-1 }{(4f/7) \, (\mathbf{B}\times\mathbf{\hat{k}})^2-1} } \; ,
\label{Ren1}
\\
\Im[n_1^{\prime}] &=& \frac{1}{\sqrt{2}} \sqrt{ \dfrac{ \sqrt{1+ \frac{\sigma^2}{\omega^2}}+1 }{(4f/7) \, (\mathbf{B}\times\mathbf{\hat{k}})^2-1 } }  \; .
\label{Imn1}
\end{eqnarray}
\end{subequations}
Consequently, these results show that the wave absorption, when $|{\bf B}\times\hat{{\bf k}}| > 57.51 \, \mbox{MeV}^{2}$,
is greater in relation to the first case of $|{\bf B}\times\hat{{\bf k}}| < 57.51 \, \mbox{MeV}^{2}$. If we convert to Tesla unit,
this magnetic field has the magnitude $|{\bf B}\times\hat{{\bf k}}| = 57.51 \, \mbox{MeV}^{2}=8.45 \times 10^{10} \, \mbox{T}$
that is in the range of magnetic fields in neutron stars \cite{Andreas1}. Notice that this magnitude is above the Schwinger's
critical field previously cited in the section \ref{sec2}.
\begin{figure*}[th]
%\vspace{-5pt}
\centering
\includegraphics[width=0.31\textwidth]{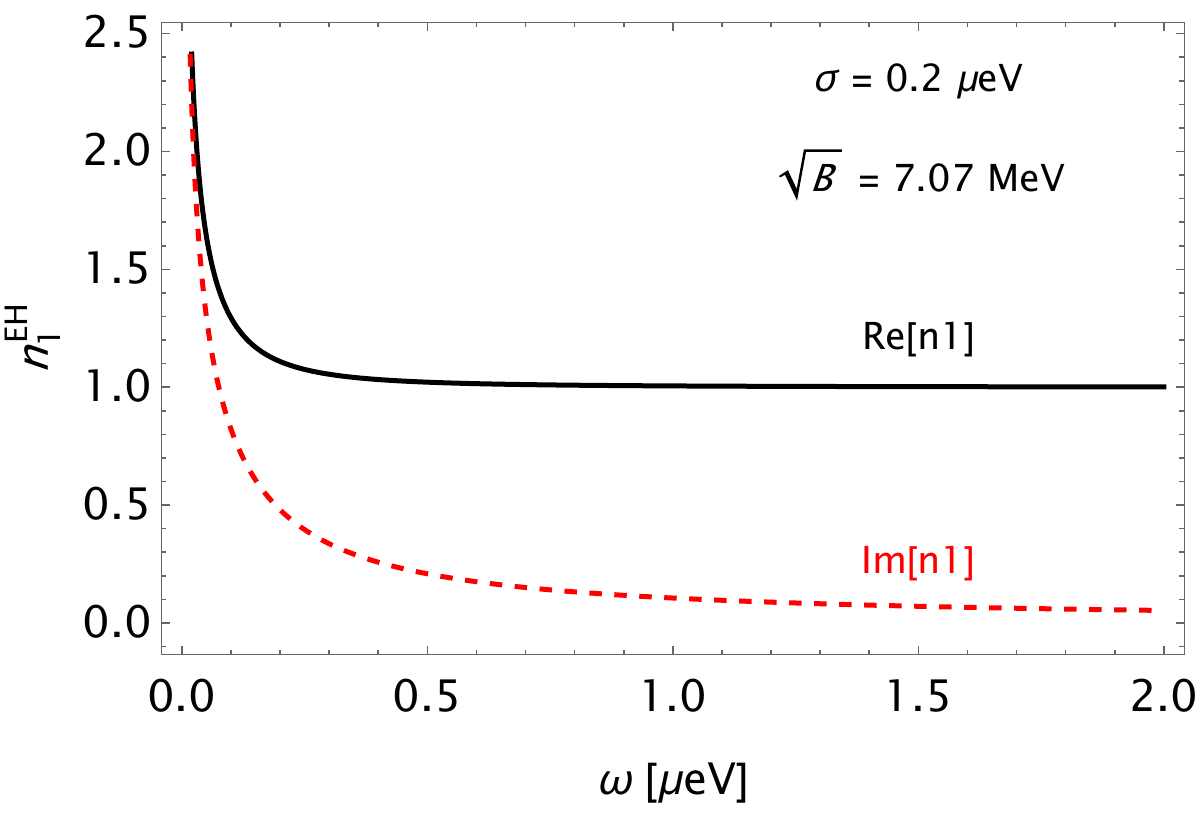}
\quad
\includegraphics[width=0.31\textwidth]{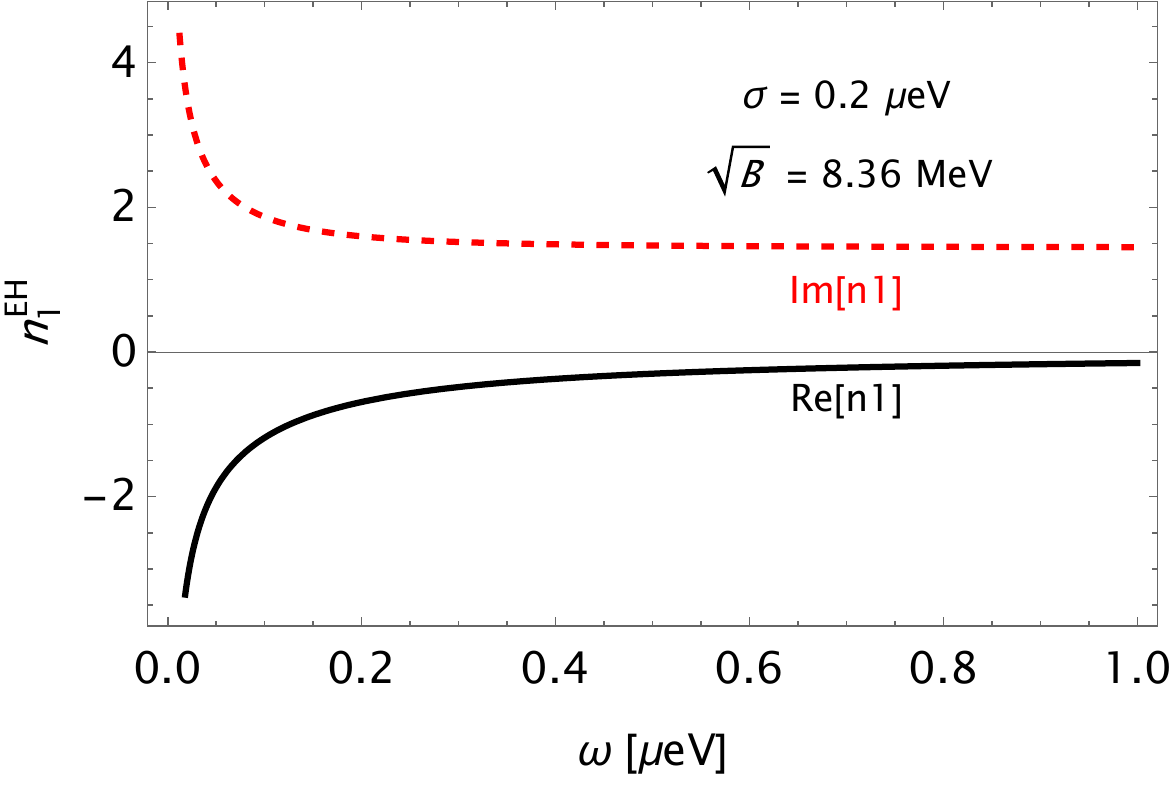}
\quad
\includegraphics[width=0.31\textwidth]{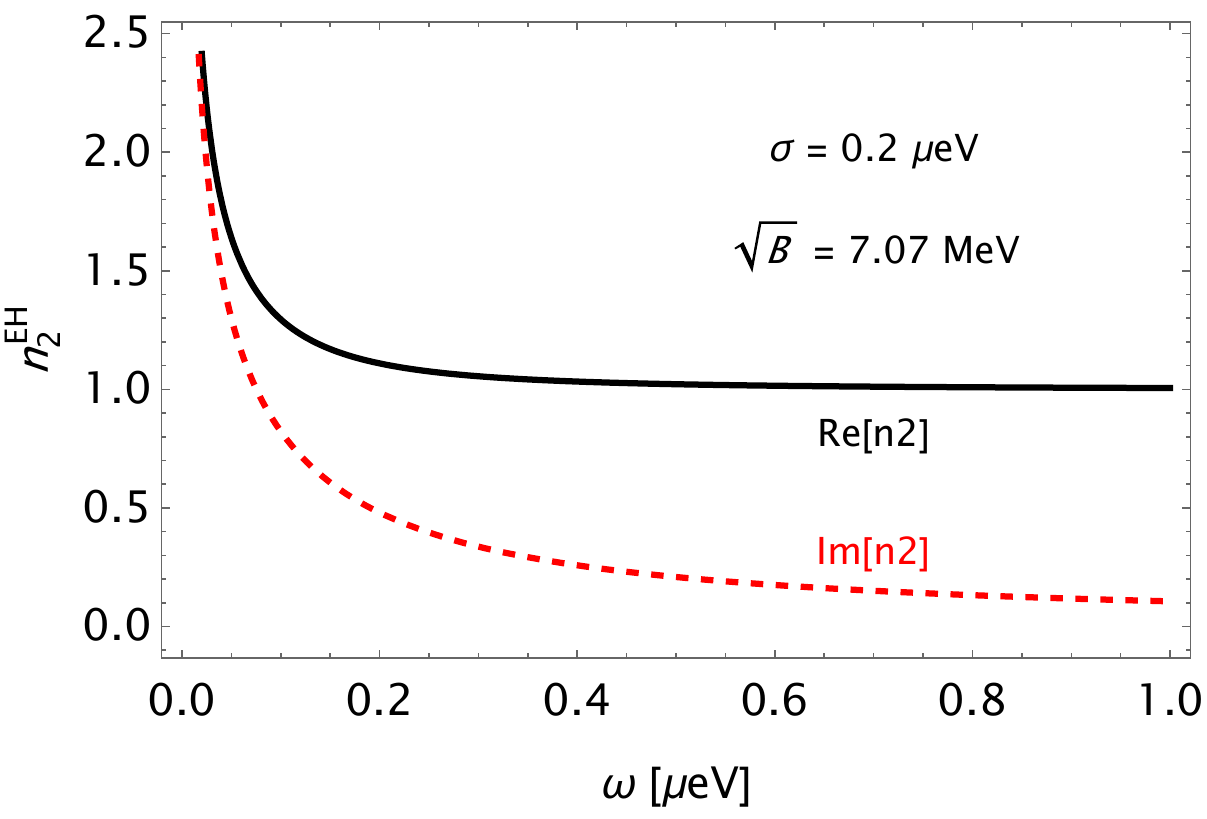}
\caption{ Left panel : The real (black line) and imaginary (red dashed line) parts of the solution $n_1$ as functions of the $\omega$-frequency, when ${\bf B}$ is perpendicular to $\hat{{\bf k}}$ for $\sqrt{|{\bf B}|}=7.07$ MeV. Middle panel : The real (black line) and imaginary (red dashed line) parts of the solution $n_1$ as functions of the $\omega$-frequency, when ${\bf B}$ is perpendicular to $\hat{{\bf k}}$ for $\sqrt{|{\bf B}|}=8.36$ MeV. Right panel : The real (black line) and imaginary (red dashed line) parts of the solution $n_2$ as functions of the $\omega$-frequency, when ${\bf B}$ is parallel to $\hat{{\bf k}}$ for $\sqrt{|{\bf B}|}=7.07$ MeV. In all these plots, we use $\sigma=0.2\,\mu\mbox{eV}$. }
\label{Ren1Imn1}
\end{figure*}
The solution of (\ref{n2}) is
%De forma análoga, desenvolve-se o $\mathbf{n}_2$, realizando a distributiva e isolando o $\mathbf{n}_2$,
%	
\begin{equation}
n_2 =\sqrt{ \dfrac{(1+i\sigma/\omega)(1+i\sigma/\omega+f\,{\bf B}^2)}{1+i\sigma/\omega + f(\mathbf{B} \cdot \mathbf{\hat{k}})^2} } \; ,
\end{equation}
that has the real and imaginary parts
\begin{subequations}
\begin{eqnarray}
\Re[n_2] &=& \frac{1}{\sqrt{2}} \, \sqrt{ \sqrt{a^2 + b^2} + a } \; ,
\label{Ren2}
\\
\Im[n_2] &=& \frac{1}{\sqrt{2}} \, \sqrt{ \sqrt{a^2 + b^2} - a } \; ,
\label{Imn2}
\end{eqnarray}
\end{subequations}
where $a$ e $b$ are defined by
\begin{widetext}
\begin{subequations}
\begin{eqnarray}
a &=& \dfrac{ 1 + \frac{\sigma^2}{\omega^2}\,[\,1+f({\bf B}\times\hat{{\bf k}})^2\,]+f\,{\bf B}^2[\,1+(\hat{{\bf B}}\cdot\hat{{\bf k}})^2+f({\bf B}\cdot\hat{{\bf k}})^2\,] }{ [ \, 1+f\,(\mathbf{B} \cdot \mathbf{\hat{k}})^2 \, ]^2 + \sigma^2/\omega^2} \; ,
\\
b &=& \frac{\sigma}{\omega} \, \dfrac{1 + \frac{\sigma^2}{\omega^2}-f\,({\bf B}\times\hat{{\bf k}})^2 + f\,{\bf B}^2[\,1+(\hat{{\bf B}}\cdot\hat{{\bf k}})^2+f({\bf B}\cdot\hat{{\bf k}})^2\,] }{ [\, 1+f\,(\mathbf{B} \cdot \mathbf{\hat{k}})^2 \, ]^2 + \sigma^2/\omega^2  } \; .
\end{eqnarray}
\end{subequations}
\end{widetext}
The real parts (\ref{Ren1}) and (\ref{Ren2}) contain the solutions of the dispersion relations,
the wavelength, and the group velocity of the plane wave. The imaginary parts (\ref{Imn1}) and (\ref{Imn2})
define the wave penetration in a conductor medium as $D=(\Im[n_i])^{-1} \, (i=1,2)$.
%
%
%Reunindo todos os resultados, obtemos as soluções de $n_1$ e $n_2$, respectivamente,
%	
%\begin{subequations}
%\begin{eqnarray} \label{n1 complet}
%\hspace{-3cm}
%n_1 =  \sqrt{ \dfrac{\sqrt{1+ \frac{\sigma^2}{\omega^2}}+1}{2 - 2(\beta^2+\mathbf{B}^2)^{-1} \, (\mathbf{B} \times \mathbf{\hat{k}})^2} } \nonumber
%\\
%+ i \;  \sqrt{ \dfrac{\sqrt{1+ \frac{\sigma^2}{\omega^2}}-1}{2 - 2(\beta^2+\mathbf{B}^2)^{-1} \, (\mathbf{B} \times \mathbf{\hat{k}})^2} } \; , \hspace{1cm}
%\\
%\hspace{-4cm}
%n_2 = \! \sqrt{ \frac{|a|}{2} } \, \sqrt{ \sqrt{1 + \frac{b^2}{a^2}} + \mbox{sign}(a) }  \nonumber
%\\
%+  i \; \sqrt{ \frac{|a|}{2} } \, \sqrt{ \sqrt{1 + \frac{b^2}{a^2}} - \mbox{sign}(a) } \; .
%\hspace{1.0 cm} \label{n2 compret}
%\end{eqnarray}
%\end{subequations}	
%	
%\begin{figure}[t]
%\vspace{-5pt}
%\centering
%\includegraphics[width=0.47\textwidth]{n2omega.pdf}
%\caption{ As partes real (linha contínua) e imaginária (linha vermelha pontilhada) da solução $n_2$ como funções da frequência $\omega$. }
%\label{Ren2Imn2}
%\end{figure}
%
The limit $f \rightarrow 0$ in (\ref{Ren1})-(\ref{Imn1}), and in (\ref{Ren2})-(\ref{Imn2}), recovers the known results of the Maxwell ED :
\begin{subequations}
\begin{eqnarray}
\lim_{f \rightarrow 0 }\Re[n_1]=\Re[n_2]=\sqrt{\sqrt{\frac{1}{4}+\frac{\sigma^2}{4\omega^2}}+\frac{1}{2} } \; ,
\\
\lim_{f \rightarrow 0 }\Im[n_1]=\Im[n_2]= \sqrt{\sqrt{\frac{1}{4}+\frac{\sigma^2}{4\omega^2}}-\frac{1}{2} } \; .
\end{eqnarray}
\end{subequations}
The results (\ref{Ren1})-(\ref{Imn1}) and (\ref{Ren2})-(\ref{Imn2}) show the dependence of the real and imaginary parts with
the angle $(\theta)$ that the magnetic background does with the wave propagation direction, {\it i.e.}, $\cos\theta=\hat{{\bf k}}\cdot\hat{{\bf B}}$. 
In the case of ${\bf k}$ parallel to ${\bf B}$, the contribution of the EH $f$-parameter disappears in the results (\ref{Ren1B}) and $(\ref{Imn1B})$. 
When ${\bf B}$ is perpendicular to $\hat{{\bf k}}$, we obtain the results
\begin{subequations}
\begin{eqnarray}
\left. \Re[n_1] \right|_{{\bf B} \perp {\bf k}} &=& \frac{1}{\sqrt{2}} \sqrt{ \dfrac{\sqrt{1+ \frac{\sigma^2}{\omega^2}}+1 }{ 1-4fB^2/7} } \; ,
\label{Ren1perp}
\\
\left. \Im[n_1] \right|_{{\bf B} \perp {\bf k}} &=& \frac{1}{\sqrt{2}} \sqrt{ \dfrac{\sqrt{1+ \frac{\sigma^2}{\omega^2}}-1 }{1-4fB^2/7 } } \; ,
\label{Imn1perp}
\\
\left. \Re[n_2] \right|_{{\bf B} \perp {\bf k}} &=& \frac{1}{\sqrt{2}} \, \sqrt{ \sqrt{(1+fB^2)^2 + \frac{\sigma^2}{\omega^2}} + 1+fB^2 } \; ,
\hspace{0.9cm}
\label{Ren2perp}
\\
\left. \Im[n_2] \right|_{{\bf B} \perp {\bf k}} &=& \frac{1}{\sqrt{2}} \, \sqrt{ \sqrt{(1+fB^2)^2 + \frac{\sigma^2}{\omega^2}} - 1-fB^2 } \; .
\label{Imn2perp}
\hspace{0.9cm}
\end{eqnarray}
\end{subequations}
%
%
%Under an intense magnetic field, the condition $|{\bf B}| \gg \beta$ reduces the results to expressions
%
%\begin{subequations}
%\begin{eqnarray}
%\Re[n_1]=\Re[n_2]=\frac{1}{|\cos\theta|} \, \sqrt{\sqrt{\frac{1}{4}+\frac{\sigma^2}{4\omega^2}}+\frac{1}{2} } \; ,
%\\
%\Im[n_1]=\Im[n_2]= \frac{1}{|\cos\theta|} \, \sqrt{\sqrt{\frac{1}{4}+\frac{\sigma^2}{4\omega^2}}-\frac{1}{2} } \; .
%\end{eqnarray}
%\end{subequations}
%
%These particular cases show that when the magnetic field is intense, the results of the Maxwell ED in a conductor are modified by the
%$\theta$-angle, and it does not depend on the magnetic field magnitude.
%

%Note que, nesses resultados, as partes real e imaginária dos índices de refração depende do ângulo em o campo magnético externo
%$\mathbf{B}$ faz com a direção de propagação de onda $\mathbf{\hat{k}}$. Na situação em que $\mathbf{B}$ é paralelo a
%$\mathbf{\hat{k}}$, a solução de $n_1$ reduz-se ao caso usual da ED de Maxwell, enquanto que $n_2$ permanece com a contribuição
%do campo magnético externo, e também do campo crítico $\beta$ de BI.
%

%
We illustrate the real and imaginary parts of $n_1$ (left panel) and $n_2$ (right panel) as functions of the $\omega$-frequency in the fig. (\ref{Ren1Imn1}). 
The left panel in (\ref{Ren1Imn1}) set the real (black line) and imaginary (red dashed line) parts of (\ref{Ren1B}) and (\ref{Imn1B}) when ${\bf B}\cdot{\bf k}=0$ 
for a magnetic field of $\sqrt{B}=7.07 \, \mbox{MeV}$, that satisfies the condition $|{\bf B}\times\hat{{\bf k}}| < 57.51 \, \mbox{MeV}^{2}$, and $\sigma=0.2\, \mu\mbox{eV}$ in natural units \footnote{In natural units, the electric resistivity has the conversion $1\Omega\cdot m=2.95 \times 10^{23} \, \mbox{GeV}^{-1}$. Therefore, electrical conductivity has energy dimension.}. When the magnetic field is perpendicular to the wave propagation direction, the $\Im[n_1]$ part decays for high frequencies in relation to $\Re[n_1]$, {\it i. e.}, the wave dispersion and absorption go to zero in the high-energy limit. The middle panel set the solutions (\ref{Ren1}) and (\ref{Imn2}) for a magnetic field of $\sqrt{B}=8.36 \, \mbox{MeV}$, that satisfies $|{\bf B}\times\hat{{\bf k}}| > 57.51 \, \mbox{MeV}^{2}$. The black line sets the real part $\Re[n_1^{\prime}]$ that is negative for any frequency. For high frequencies, the real part is null, and as consequence, the wave absorption is total. The right panel in (\ref{Ren1Imn1}) illustrates the real and imaginary parts of the $n_{2}$-solution as functions of the $\omega$-frequency. In this case, we choose the magnetic field parallel to the direction of the wave propagation direction. The dispersion falls down faster in relation to wave absorption for high frequencies. 

%A figura (\ref{Ren1Imn1}) mostra que $\Re[n_1]$ cai mais rapidamente com a frequência do que $\Im[n_1]$, ou seja, a dispersão e a absorção da onda vão a zero no regime de altas frequências. Na figura (\ref{Ren2Imn2}), as partes real e imaginária da solução $n_2$ são ilustradas como funções da frequência $\omega$ para o caso do campo magnético paralelo a direção de propagação de onda. A segunda solução tem um comportamento similar ao da solução de $n_1$.
%
%\begin{figure}[t]
%\vspace{-5pt}
%\centering
%\includegraphics[width=0.85\textwidth]{n2omega.pdf}
%\caption{ As partes real (linha preta) e imaginária (linha vermelha pontilhada) da solução $n_2$ como funções da frequência $\omega$. }
%\label{Ren2Imn2}
%\end{figure}
%

%
%%%%%%%%%%%%%%%%%%%
%

\section{The wave dispersion in the presence of an magnetic current density}
\label{sec4}
In this section, we study the dispersion effects for the case of a magnetic conductivity current. The nature of this current density is associated with the chirality between left- and right-handed fermions when it is submitted to an external magnetic field \cite{Burkov}. From the classical point of view, the magnetic current density is
given by
\begin{eqnarray}\label{JB}
\mathbf{J} = \sigma_{b} \, \mathbf{B}_0 \; ,
\end{eqnarray}
where $\sigma_{b}$ is the magnetic conductivity, that we consider isotropic throughout the material medium. Using the prescription of the magnetic background, in which $\mathbf{B}_0=\mathbf{b} + \mathbf{B}$, the linearized equations in the presence of the current density (\ref{JB}) are read below
%
%espaço dos momentos, para uma corrente magnética no condutor, $\mathbf{J} = \sigma \,(\mathbf{b} + \mathbf{B})$, onde $\sigma$ é a condutibilidade magnética, que vária para cada material,
%	
\begin{subequations} \label{eq.linea.b}
\begin{eqnarray}
\mathbf{k} \cdot {\mathbf{e_0}} + f \; (\mathbf{B} \cdot \mathbf{k}) (\mathbf{B} \cdot \mathbf{e_0}) \!&=& \rho_{0}(k) \, ,
\\
\mathbf{k} \times {\mathbf{e_0}}  &=& \omega \, \mathbf{b_0} \; ,
\label{eqd rot e.b}
%\hspace{0.2cm} , \hspace{0.2cm}
\\
\mathbf{k} \cdot {\mathbf{b_0}} &=& 0 \,
\hspace{0.2cm} , \hspace{0.2cm}
\\
 \mathbf{k} \times {\mathbf{b_0}} + \frac{4f}{7}\, (\mathbf{B} \times \mathbf{k}) (\mathbf{B} \cdot {\mathbf{b_0}}) &=& - i \, \sigma_b \,\mathbf{b_0} \nonumber
 \\
 - i \, \sigma_b \, \mathbf{B} \, \delta^3(\mathbf{k})\, \delta(\omega)
 - f \, \omega \, \mathbf{B} \, (\mathbf{B} \cdot \mathbf{e_0}) -\omega \, \mathbf{e_0} \; ,
 \hspace{-1.7cm}
\label{eqdivB.b}
\end{eqnarray}
\end{subequations}
where we have substituted the plane wave solutions via Fourier transform for the propagating EM fields $\mathbf{e}$, $\mathbf{b}$ and the charge density $\rho$. Combining the Faraday law with the eq. (\ref{eqdivB.b}), we obtain the wave equation for the electric amplitude $e_{0j}$ :
\begin{eqnarray}\label{EqmatrixO}
O_{i j} \,e_{0j} = 0 \; ,
\end{eqnarray}
where the matrix elements of $O_{ij}$ are read below
\begin{eqnarray}\label{Oij}
O_{i j} &=& \left( 1 - \mathbf{n}^2\right)\, \delta_{ij}+n_i \, n_j
- i \, \frac{\sigma_b}{\omega}  \, \epsilon_{i j k}\, n_k
\nonumber \\
\hspace{-0.5cm}
&&
+ \frac{4f}{7}\, (\mathbf{B} \times \mathbf{n})_i \, (\mathbf{B} \times \mathbf{n})_j
+ f \, B_i \, B_j \; .
\end{eqnarray}
%
%\begin{eqnarray}
%	\hspace{-1.0 cm} \mathbf{n} \cdot (\mathbf{n} \cdot \mathbf{e_0}) + \mathbf{e_0} ( 1 - \mathbf{n}^2 ) + ({\mathbf{B}}^2+\beta^2)^{-1} \; (\mathbf{B} \times \mathbf{n}) (\mathbf{B} \times \mathbf{n}) \cdot \mathbf{e_0} + i\, \frac{\sigma}{\omega}  \,(\mathbf{n} \times {\mathbf{e_0}}) + \beta^2 \, \mathbf{B} \, (\mathbf{B} \cdot \mathbf{e_0}) \!\!\!&=&\!\!\! 0 \; . \hspace{1.0 cm}
%	\end{eqnarray}
%	
%	Transformando para componentes, por definição, o produto vetorial em componentes é $(\mathbf{n} \times {\mathbf{e_0}})_i := - \, \varepsilon_{i\, j \, k} \, e_j \, n_k$, logo,
%	
%	\begin{eqnarray}
%	\hspace{-1.0 cm} [\mathbf{n}_i \,\mathbf{n}_j + ( 1 - \mathbf{n}^2) \, \delta_{i\,j} + ({\mathbf{B}}^2+\beta^2)^{-1} \; (\mathbf{B} \times \mathbf{n})_i (\mathbf{B} \times \mathbf{n})_j - i\, \frac{\sigma}{\omega}  \, \varepsilon_{i\, j \, k}\, n_k + \beta^2 \, \mathbf{B}_i \, \mathbf{B}_j] \, \mathbf{e}_j  \!\!\!&=&\!\!\! 0 \; . \hspace{1.0 cm}
%	\end{eqnarray}
%	
%	A equação $M_{i \,j} \,{\mathbf{e_0}}_j = 0$, para ondas planas ser solução, a matriz $M_{i \, j}$,
%
The non-trivial solution of (\ref{EqmatrixO}) requires the condition $\det(O_{ij})=0$,
that yields the equation
\begin{eqnarray}
(1+f{\bf B}^2)\left[1+n^4-n^2\left(2+\frac{\sigma_{b}^2}{\omega^2}\right) \right]
\nonumber \\
+n^2\left[ \frac{11}{7}(1-n^2)+\frac{4}{7}\,f{\bf B}^2+\frac{\sigma_b^2}{\omega^2} \right]f({\bf B}\times\hat{{\bf k}})^2
\nonumber \\
-\frac{4}{7}\,n^4f^2({\bf B}\times\hat{{\bf k}})^2\,({\bf B}\cdot\hat{{\bf k}})^2
=0 \; .
\end{eqnarray}
The correspondent solutions are given by	
%	
%	
%	precisa-se que o determinante seja igual a zero. Com isso, surgirá quatro soluções distintas, escolhendo os índices de refração positivos, tem-se duas equações,
%	
	%que será utilizada as com índices de refração positivos. Então, a solução será dois índices de refração diferentes,
%	
	%\begin{subequations}
	%\begin{eqnarray}
	%\mathbf{n}_1 \!\!\!&=&\!\!\! \sqrt{\frac{(\beta ^2 + \mathbf{B}^2 )\, \left( 2 (\mathbf{B}^2 - (\mathbf{B} \times \hat{k})^2)\, (2 + \, \epsilon ^2)  + \beta ^2 \left( 4 + 2 \, \epsilon ^2  - \sqrt{\frac{\epsilon ^2 (4 + \epsilon ^2) 4 ( \mathbf{B}^2 + \,\beta^2 - ( \mathbf{B} \times \hat{k})^2)^2}{\beta ^4}} \right) \right)}{ 4 ( \,\mathbf{B}^2 +  \beta ^2 - (\mathbf{B} \times \hat{k})^2)^2}} \, ,
	%\\  \nonumber
	%\\
	%\mathbf{n}_2 \!\!\!&=&\!\!\! \sqrt{\frac{(\beta ^2 + \mathbf{B}^2 )\, (2 (\mathbf{B}^2 - (\mathbf{B} \times \hat{k})^2)\, (2 + \epsilon ^2)  + \beta ^2 \left( 4 + 2 \, \epsilon ^2  + \sqrt{\frac{\epsilon ^2 (4 + \epsilon ^2) 4 ( \mathbf{B}^2 + \,\beta^2 - ( \mathbf{B} \times \hat{k})^2))^2}{\beta ^4}} \right)}{ 4 (\mathbf{B}^2+ 2 \beta ^2 -  (\mathbf{B} \times \hat{k})^2)^2}} \, .
	%\end{eqnarray}
	%\end{subequations}
%	
%Realizando algumas simplificações, temos
%	
%
\begin{figure*}[th]
%\vspace{-5pt}
\centering
\includegraphics[width=0.45\textwidth]{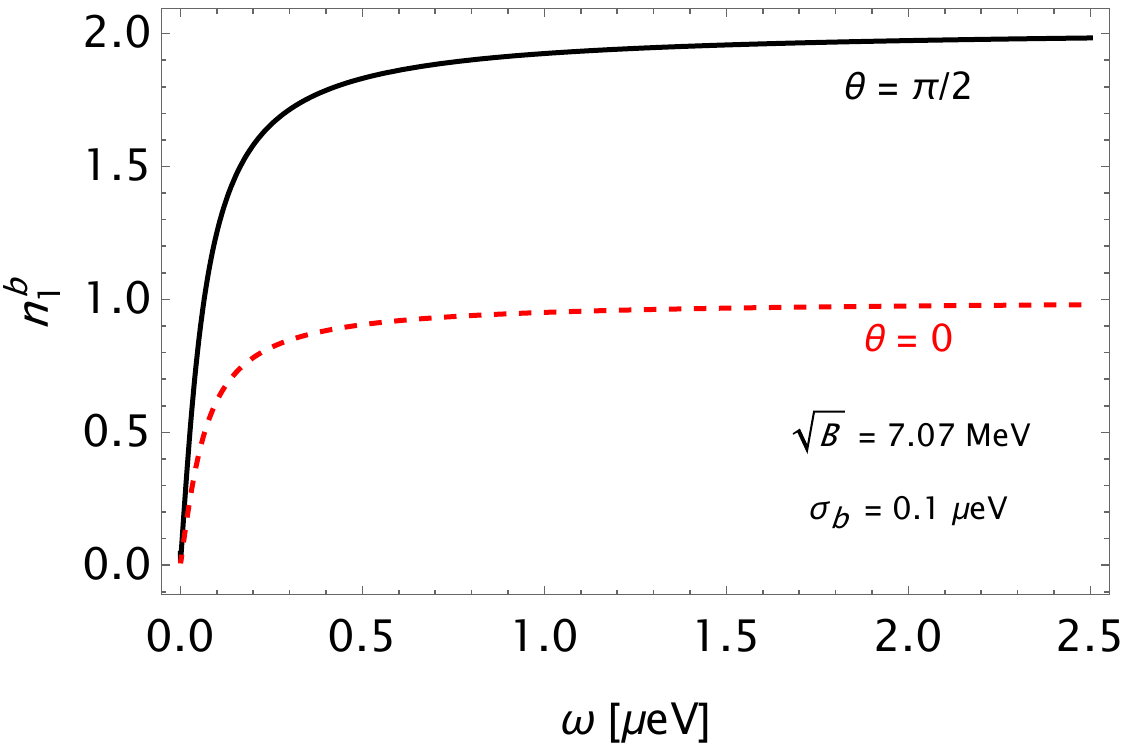}
\quad \quad
\includegraphics[width=0.45\textwidth]{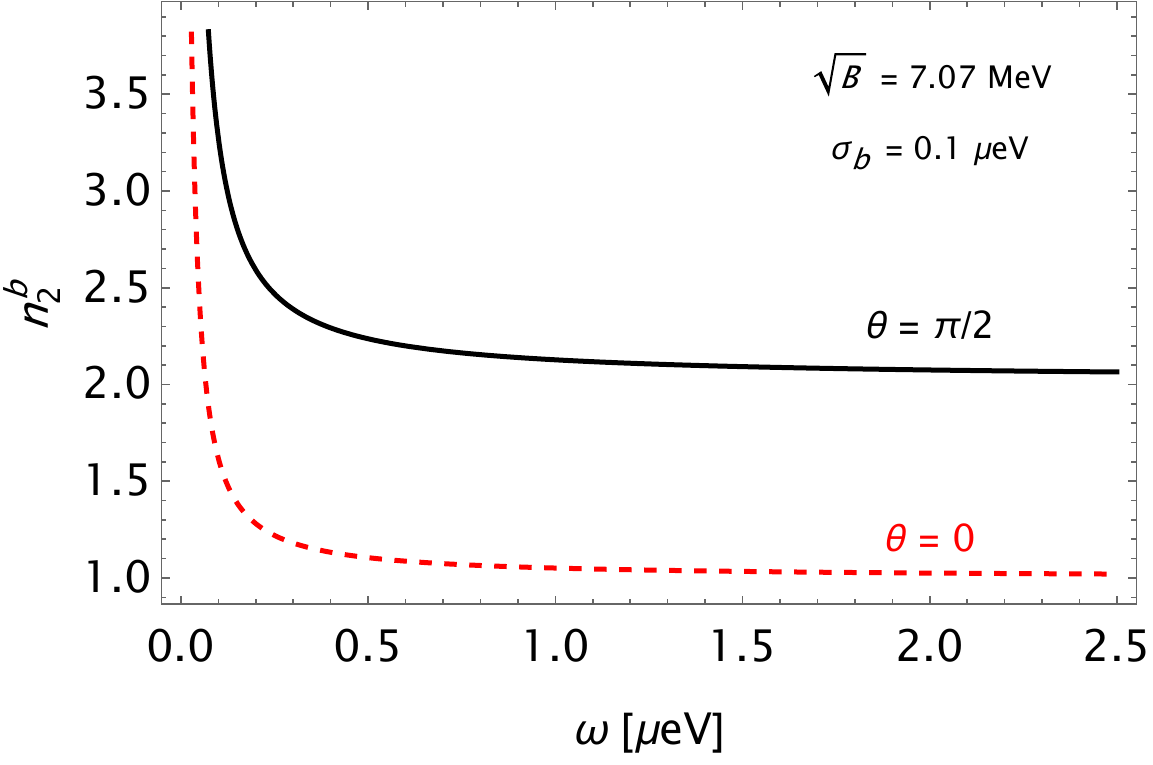}
\caption{ Left panel : The $n_1$-solution from (\ref{n1 b}) as function of the $\omega$-frequency. Right panel : The $n_2$-solution from (\ref{n1 b})
as function of the $\omega$-frequency. In both plots, the solid black line means the case of ${\bf B}\cdot {\bf k}=0$ (perpendiculars), whereas the dashed 
red lines set the case of ${\bf B} \times {\bf k}=0$ (parallels). We choose $\sqrt{|{\bf B}|}=7.07$ MeV and $\sigma_{b}=0.1 \, \mu\mbox{eV}$ in both plots. }
\label{n1B}
\end{figure*}
%
%{\color{red}
%\begin{subequations}
\begin{eqnarray}
n_1^b = \sqrt{ \frac{N_1}{D} }
%\label{n1 b}
\hspace{0.4cm} \mbox{and} \hspace{0.4cm}
n_2^b = \sqrt{ \frac{N_2}{D} } \; , \hspace{1cm}
\label{n1 b}
\end{eqnarray}
%\end{subequations}
%
where
\begin{widetext}
\begin{eqnarray}
N_1 &=& (1+f{\bf B}^2)\left(2+\frac{\sigma_b^2}{\omega^2}\right)-\left(\frac{11}{7}+\frac{4}{7}\,f\,{\bf B}^2+\frac{\sigma_b^2}{\omega^2}\right)f({\bf B}\times\hat{{\bf k}})^2
\nonumber \\
&&
\hspace{-0.5cm}
-\left\{ \frac{\sigma_{b}^2}{\omega^2}\,(1+f\,{\bf B}^2)\left(4+\frac{\sigma_b^2}{\omega^2} \right)-2f({\bf B}\times\hat{{\bf k}})^2\,\frac{\sigma_b^2}{\omega^2}\left( \frac{25}{7}+\frac{4}{7}\,f\,{\bf B}^2+\frac{\sigma_b^2}{\omega^2} \right)
\right.
\nonumber \\
&&
\hspace{-0.5cm}
\left.
+f^2({\bf B}\times\hat{{\bf k}})^4\left[ \left( \frac{3}{7}-\frac{4}{7}\,f\,{\bf B}^2 \right)^2+\left( \frac{11}{7}+\frac{4}{7}\,f\,{\bf B}^2  \right)2\,\frac{\sigma_b^2}{\omega^2}+\frac{\sigma_b^4}{\omega^4} \right]
\right\}^{1/2} \; ,
\\
N_2 &=&  (1+f{\bf B}^2)\left(2+\frac{\sigma_b^2}{\omega^2}\right)-\left(\frac{11}{7}+\frac{4}{7}\,f\,{\bf B}^2+\frac{\sigma_b^2}{\omega^2}\right)f({\bf B}\times\hat{{\bf k}})^2
\nonumber \\
&&
\hspace{-0.5cm}
+\left\{ \frac{\sigma_{b}^2}{\omega^2}\,(1+f\,{\bf B}^2)\left(4+\frac{\sigma_b^2}{\omega^2} \right)-2f({\bf B}\times\hat{{\bf k}})^2\,\frac{\sigma_b^2}{\omega^2}\left( \frac{25}{7}+\frac{4}{7}\,f\,{\bf B}^2+\frac{\sigma_b^2}{\omega^2} \right)
\right.
\nonumber \\
&&
\hspace{-0.5cm}
\left.
+f^2({\bf B}\times\hat{{\bf k}})^4\left[ \left( \frac{3}{7}-\frac{4}{7}\,f\,{\bf B}^2 \right)^2+\left( \frac{11}{7}+\frac{4}{7}\,f\,{\bf B}^2  \right)2\,\frac{\sigma_b^2}{\omega^2}+\frac{\sigma_b^4}{\omega^4} \right]
\right\}^{1/2} \; ,
\\
D &=& 2(1+f\,{\bf B}^2)-\frac{8}{7}\,f({\bf B}\times\hat{{\bf k}})^2\left( \frac{11}{4}+f\,{\bf B}^2 \right)
+\frac{8}{7}\,f^2({\bf B}\times\hat{{\bf k}})^4 \; .
\end{eqnarray}
\end{widetext}
%
%}
%
%that also imposes the condition (\ref{condB}).
%
Notice that, in both solutions, there is no absorption in this case of an isotropic magnetic conductivity.
In $n_{1}^{b}$ and $n_{2}^{b}$, also emerge the dependence on the $\theta$-angle that $\mathbf{B}$ does with $\mathbf{\hat{k}}$-direction.
The limit $f \rightarrow 0$ recovers the known result in the literature \cite{Pedro}
\begin{subequations}
\begin{eqnarray}\label{n1bf=0}
\lim_{f \rightarrow 0 } n_1^{b} &=& 1+\frac{\sigma_{b}^2}{2\omega^2}-\frac{\sigma_{b}}{\omega}\sqrt{1+\frac{\sigma_b^2}{4\omega^2} } \; ,
\label{n1bf=0}
\\
\lim_{f \rightarrow 0 } n_2^{b} &=& 1+\frac{\sigma_{b}^2}{2\omega^2}+\frac{\sigma_{b}}{\omega}\sqrt{1+\frac{\sigma_b^2}{4\omega^2} } \; .
\label{n2bf=0}
\end{eqnarray}
\end{subequations}

In the case of $\mathbf{B}$ parallel to $\mathbf{\hat{k}}$, the non-linear contribution of the $f$-parameter is canceled in
(\ref{n1 b}), that so reduce to (\ref{n1bf=0}) and (\ref{n2bf=0}). For $\mathbf{B}$ perpendicular to $\mathbf{\hat{k}}$, the $f$-parameter and
the magnetic background contribute to the solutions, such that the results are
\begin{widetext}
\begin{subequations}
\begin{eqnarray}
\left. n_1^b\right|_{{\bf B}\perp\hat{{\bf k}}} &=& \sqrt{ \frac{14+3fB^2-4f^2B^4+7\sigma_b^2/\omega^2}{14-8fB^2}
-\frac{1}{14-8fB^2}\sqrt{ 196 \, \frac{\sigma_b^2}{\omega^2}+\left[ fB^2(3-4\,fB^2)+\frac{\sigma_b^2}{\omega^2} \right]^2 } } \; ,
\label{n1 bperp}
\\
%\hspace{0.4cm} \mbox{and} \hspace{0.4cm}
\left. n_2^b\right|_{{\bf B}\perp\hat{{\bf k}}} &=& \sqrt{ \frac{14+3fB^2-4f^2B^4+7\sigma_b^2/\omega^2}{14-8fB^2}
+\frac{1}{14-8fB^2}\sqrt{ 196 \, \frac{\sigma_b^2}{\omega^2}+\left[ fB^2(3-4\,fB^2)+\frac{\sigma_b^2}{\omega^2} \right]^2 } } \; .
\hspace{1cm}
\label{n2 bperp}
\end{eqnarray}
\end{subequations}
\end{widetext}
The $n_1$-solution, when $\mathbf{B}\cdot\mathbf{k}=0$ (black line),
and when $\mathbf{B} \times \mathbf{k}={\bf 0}$ (red line) are both showed in the fig. (\ref{n1B}).
In this plot, we choose $\sqrt{|{\bf B}|}=7.07 \, \mbox{MeV}$, and $\sigma_b=0.1 \, \mu\mbox{eV}$. In high frequency range, the curves goes to a maximum refractive index, that satisfies the condition  $n_{1\perp}>n_{1\parallel}$, where $n_{1\perp} \simeq 2.023$, and $n_{1\parallel}\simeq 1.0$, when $\omega\rightarrow \infty$.

%	
%Observe os índices de refração são bem parecidos, o que os diferem é apenas um sinal negativo. Com isso, pode-se realizar o plot da equação (\ref{n1 b}), de $\mathbf{n}_1$ em função da velocidade angular ($\omega$).  Como os índices de refração dependem do ângulo entre o campo magnético externo ($\mathbf{B}$) e o vetor unitário de propagação de onda ($\hat{k}$), será realizado o plot para dois casos, o paralelo e o perpendicular.
	
%
%
%$\mathbf{n}_1$ e $\mathbf{n}_2$, com o índice de refração em função da velocidade angular da onda ($\omega$).
%
%
%Analisando ambos os casos, tem-se que o índice de refração satura em valores diferentes, dependendo do ângulo entre $\mathbf{B}$ e $\hat{k}$.
%Observe, que os índices de refração dependem do ângulo entre $\mathbf{B}$ e $\hat{k}$. Porém, são bem parecidos, o que os diferenciam é apenas um sinal de menos. Com isso, pode-se realizar o plot da equação (\ref{n1 b}), do índice de refração $\mathbf{n}_1$, para o caso perpendicular e paralelo, em função da velocidade angular.
%	
	%De forma analoga, para $n2$,
%	
	%\begin{eqnarray}
	%\mathbf{n}_2 = \sqrt{\frac{(\beta ^2 + \mathbf{B}^2 ) \, \left( ( 2 + \, \epsilon ^2) + \sqrt{\epsilon ^2 (4 + \epsilon ^2)}  \right)}{ 2 ( \,\mathbf{B}^2 +  \beta ^2 - (\mathbf{B} \times \hat{k})^2)}} \, .
	%\end{eqnarray}
%
%\begin{figure}[t]
%\vspace{-5pt}
%\centering
%\includegraphics[width=0.85\textwidth]{N1B.pdf}
%\caption{ A solução $n_1$ de (\ref{n1 b}) como função da frequência $\omega$. }
%\label{n1B}
%\end{figure}
%
The $n_2$-solution is shown in the right panel from the fig. (\ref{n1B}). The black line corresponds to $\mathbf{B}$ perpendicular to $\mathbf{\hat{k}}$, and the dashed red line is the case of $\mathbf{B}$ parallel to $\mathbf{\hat{k}}$. In this figure, we also consider $\sqrt{|{\bf B}|}=7.07 \, \mbox{MeV}$ and $\sigma_b=0.1 \, \mu\mbox{eV}$. In the right panel, the curves have a horizontal asymptotes decay for $n_{2\perp} \simeq 2.023$ and $n_{2\parallel}\simeq 1.0$, for high frequency.
%

%
%O gráfico da solução $n_2$ está mostrado na figura (\ref{n2B}). A linha preta corresponde
%à $\mathbf{B}$ perpendicular a $\mathbf{\hat{k}}$, e a linha pontilhada vermelha é a curva
%de $\mathbf{B}$ paralelo a $\mathbf{\hat{k}}$. Nesta figura também consideramos
%$\beta=2.0 \, \mbox{MeV}^2$, $B=2.5 \, \mbox{MeV}^2$ e $\sigma=0.1 \, \mbox{MeV}$.
%Agora, as curvas atingem assíntotas horizontais mínimas para altas frequências, onde
%$n_{2\perp} \simeq 1.6$ e $n_{2\parallel}\simeq 1.0$, onde $\omega\rightarrow \infty$.
%	
%	
%\begin{figure*}[th]
%\vspace{-5pt}
%\centering
%\includegraphics[width=0.47\textwidth]{N1B.pdf}
%\quad
%\includegraphics[width=0.47\textwidth]{N2B.pdf}
%\caption{ A solução $n_2$ de (\ref{n2 b}) como função da frequência $\omega$. }
%\label{n2B}
%\end{figure*}
%\begin{figure}[t]
%\vspace{-5pt}
%\centering
%\includegraphics[width=0.85\textwidth]{N2B.pdf}
%\caption{ A solução $n_2$ de (\ref{n2 b}) como função da frequência $\omega$. }
%\label{n2B}
%\end{figure}
%
%Para a figura 4, foi utilizado a solução $\mathbf{n}_2$, o índice de refração é inversamente proporcional a $\omega$. É possível observar, quando $\mathbf{B}$ é perpendicular a $\hat{k}$, há um decaimento abrupto, diferente do caso em paralelo, onde a queda é suave. Ademais, a assíntota do gráfico perpendicular ocorre quando o índice de refração tende a um, diferente para o caso paralelo que tente a $\sqrt{5}$.
%

\section{Birefringence phenomenon}
\label{sec5}
The birefringence phenomenology is associated with the difference of the wave polarization
in relation to direction of the magnetic background field. We start this analysis considering
the magnetic background field on ${\cal Z}$-direction, ${\bf B}=B\,\hat{{\bf z}}$,
and the wave vector pointed on ${\cal X}$-direction, {\it i.e.}, ${\bf k}=k\,\hat{{\bf x}}$.
In the first case, we assume the linear wave polarization on the ${\cal Z}$-axis, that is,
${\bf e}_{0}=e_{03}\,\hat{{\bf z}}$. Thereby, we have the situation in which ${\bf B}$ is parallel
to wave amplitude, and the wave equation (\ref{EqMij}) is reduced to
\begin{eqnarray}
\left(1-n_{\parallel}^2+\frac{i\sigma}{\omega}+f\,B^{2} \right)e_{03}=0 \; ,
\end{eqnarray}
whose solution is
\begin{eqnarray}
n_{\parallel}=\sqrt{ 1+\frac{i\sigma}{\omega}+f\,B^2 } \; ,
\end{eqnarray}
where we denote the refractive index as $n_{\parallel}$. The second case is when the wave polarization
is on the ${\cal Y}$-axis, ${\bf e}_{0}=e_{02}\,\hat{{\bf y}}$, where ${\bf B}$ is now perpendicular to wave
polarization direction. Under these conditions, the eq. (\ref{EqMij}) is given by
\begin{eqnarray}
\left(1-n_{\perp}^2+\frac{i\sigma}{\omega}+\frac{4f}{7}\,B^{2}\,n_{\perp}^2 \right)e_{02}=0 \; ,
\end{eqnarray}
in which the solution is
\begin{eqnarray}
n_{\perp}=\sqrt{ \left(1+\frac{i\sigma}{\omega}\right)\left(1-\frac{4f\,B^2}{7}\right)^{\!-1}  } \; .
\end{eqnarray}
%
%
%Since the refractive index depend on the $\theta$-angle that we have defined as $\cos\theta=\mathbf{B}\cdot\mathbf{\hat{k}}$,
%the birefringence phenomenon emerges from the solutions in both electric and magnetic conductivity current cases. In the case of the
%Ohmic current, we consider the real part of the solutions for the analysis of the birefringence.
%
The birefringence is defines as the difference between
the parallel and perpendicular of the refractive indices
$\delta n = n_{\parallel}-n_{\perp}$, that is,
\begin{equation}\label{deltane}
\delta n = \sqrt{ 1+\frac{i\sigma}{\omega}+fB^2 } - \sqrt{ \left(1+\frac{i\sigma}{\omega}\right)\left(1-\frac{4fB^2}{7}\right)^{\!-1}  } \; .
\end{equation}
%
%
%where $i=\{ \, 1 \, , \, 2 \, \}$, $n_{i\perp}^{e}$ is the refractive index when $\mathbf{B}$ is perpendicular to $\mathbf{\hat{k}}$,
%and $n_{i\parallel}^{e}$ denotes the refractive index when $\mathbf{B}$ is parallel to $\mathbf{\hat{k}}$. Substituting
%the real parts of (\ref{Ren1}) and (\ref{Ren2}), we obtain
%
%\begin{subequations}
%\begin{eqnarray}
%\delta n_{1}^{e} \!&=&\! \left(\! \sqrt{1+\frac{{\bf B}^2}{\beta^2}}-1 \!\right) \! \sqrt{ \sqrt{ \frac{1}{4} + \frac{\sigma^2}{4\omega^2} } + \frac{1}{2} %} \; ,
%\label{deltan1e}
%\\
%\delta n_{2}^{e} \!&=&\! \sqrt{ \sqrt{\frac{1}{4}+\frac{{\bf B}^2}{4\beta^2}+\frac{\sigma^2}{4\omega^2}}+\frac{1}{2}+\frac{{\bf B}^2}{2\beta^2} }
%\nonumber \\
%&&
%-\sqrt{ \sqrt{\frac{1}{4}+\frac{\sigma^2}{4\omega^2}}+\frac{1}{2} } \; .
%\label{deltan2e}
%\end{eqnarray}
%\end{subequations}
%
%It is worth to highlight that the birefringence disappears $\delta n_{i}^{e}=0$, when $\beta\rightarrow \infty$.
%

\begin{figure}[t]
%\vspace{-5pt}
\centering
\includegraphics[width=0.49\textwidth]{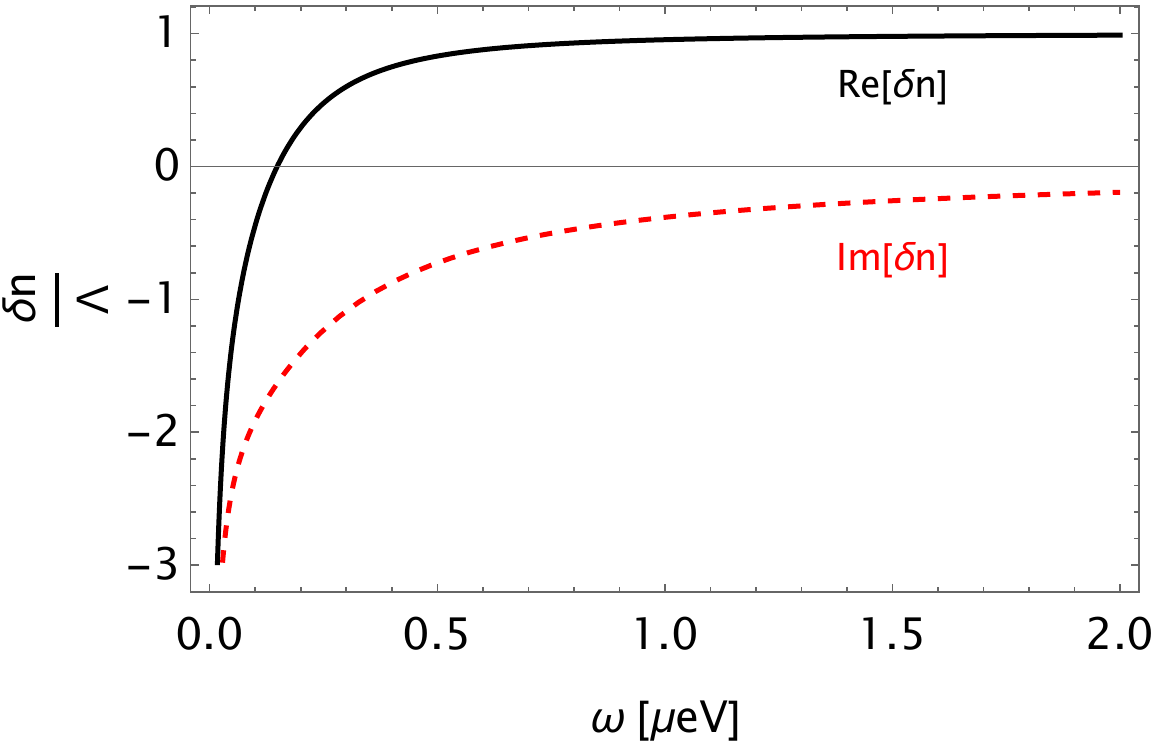}
\caption{ The birefringence curves for $\delta n$ over $\Lambda=3fB^2/14=3.27 \times 10^{-22}$ as function of the $\omega$-frequency.
The black line is the real part of $\delta n$, whereas the and dashed red line is the imaginary part. We choose $\sigma=0.21 \, \mu\mbox{eV}$.  }
\label{figBire}
\end{figure}

%
In the polarization vacuum with laser (PVLAS) experiment \cite{25years}, the magnetic field is
$|{\bf B}|=2.5\,\mbox{T}=1.7\times10^{-9}\,\mbox{MeV}^2$, in which we can consider $f\,B^2 \ll 1$
in (\ref{deltane}). Thus, the birefringence effect must be investigate in a regime of weak magnetic field. 
Considering this approximation, we obtain  :
\begin{eqnarray}
\delta n \simeq \Re[\delta n] + i \, \Im[\delta n] \; ,
\end{eqnarray}
where the real an imaginary parts are, respectively, given by
\begin{subequations}
\begin{eqnarray}
\Re[\delta n] &=& \frac{3fB^2/14}{\sqrt{1+\sigma^2/\omega^2}} \left[ \, \sqrt{ \sqrt{ \frac{1}{4} + \frac{\sigma^2}{4\omega^2} } +\frac{1}{2} }
\right.
\nonumber \\
&&
\left.
-\frac{4\sigma}{3\omega} \, \sqrt{ \sqrt{ \frac{1}{4} + \frac{\sigma^2}{4\omega^2} } - \frac{1}{2} } \, \right] \; ,
\label{Redeltan}
\\
\Im[\delta n] &=& \frac{-3fB^2/14}{\sqrt{1+\sigma^2/\omega^2}} \left[ \, \frac{4\sigma}{3\omega} \,
\sqrt{ \sqrt{ \frac{1}{4} + \frac{\sigma^2}{4\omega^2} } +\frac{1}{2} }
\right.
\nonumber \\
&&
\left.
+ \sqrt{ \sqrt{ \frac{1}{4} + \frac{\sigma^2}{4\omega^2} } - \frac{1}{2} } \, \right] \; .
\label{Imdeltan}
\end{eqnarray}
\end{subequations}
Removing the Ohm law with $\sigma \rightarrow 0$, we recover the results
$\Re[\delta n]=3fB^2/14=3.27 \times 10^{-22}$ and $\Im[\delta n]=0$, that is consistent with the PVLAS
experiment \cite{25years}. For a perfect conductor, we take $\sigma \gg \omega$, in which the results (\ref{Redeltan})
and (\ref{Imdeltan}) are reduced to
%
%\begin{subequations}
\begin{eqnarray}\label{deltanlow}
\Re[\delta n] = \Im[\delta n] \simeq -\frac{2fB^2}{7}\sqrt{ \frac{\sigma}{2\omega} } \; .
\end{eqnarray}
The birefringence curves of $\Delta n/\Lambda$ as functions of the $\omega$-frequency are shown in the figure (\ref{figBire}), in which we
define $\Lambda:=3fB^2/14=3.27 \times 10^{-22}$. The black line sets the real part $\Re[\delta n]/\Lambda$, and $\Im[\delta n]/\Lambda$ is
illustrated by the dashed red line. In this plot, we choose
%$\sqrt{\beta}=16 \, \mbox{MeV}$, $\sqrt{|\mathbf{B}|}=8.25 \, \mbox{MeV}$ and
$\sigma=0.21 \, \mu\mbox{eV}$. The range of high frequencies shows the real part at
$\Re[\delta n]=\Lambda=3.27 \times 10^{-22}$, and the imaginary part goes to zero. In low frequencies,
both curves indicate divergences at $\omega\rightarrow 0$ that are given by (\ref{deltanlow}).
%
%\begin{eqnarray}
%\Re[\delta n]=\Im[\delta n] \simeq -\frac{4\Lambda}{3}\sqrt{ \frac{\sigma}{\omega} } \; ,
%\end{eqnarray}
%
%with $\omega \ll \sigma$.
%
%
%As curvas da birrefringência como funções da frequência estão mostradas na figura (\ref{figBire}).
%Para esse gráfico, usamos $\beta=2.0 \, \mbox{MeV}^2$, $B=2.5 \, \mbox{MeV}^2$ e $\sigma=0.1 \, \mbox{MeV}$.	
%
%\vspace{0.5cm}
%
When the black line intercepts the frequency axis, the real part of $\delta n$ is null and the birefringence has a pure absorption. 
The correspondent solution is read by the frequency
\begin{eqnarray}
\omega=\frac{4\sigma}{\sqrt{33}} \; .
\end{eqnarray}

Using this same scenario of birefringence, the contribution of the magnetic conductivity $\sigma_{b}$ is null
in the component ${\cal O}_{33}$ (case in which ${\bf B}$ is parallel to ${\bf e}_{0}$), and also in ${\cal O}_{22}$ (case of
${\bf B}$ perpendicular to ${\bf e}_{0}$). Thereby, the birefringence that emerges from (\ref{Oij}) is same one
in relation to (\ref{Mij}), when the Ohm law is null.

\section{Concluding comments}
\label{sec6}
In this paper, we study the dispersion and absorption of waves in the linearized Euler-Heisenberg (EH)
electrodynamics governed by the Ohmic and magnet current densities. The linearization of the EH ED
is introduced through a propagating electromagnetic field added to a uniform and constant magnetic background field.
The EH non-linear lagrangian in expanded up to second order for small propagating effects, and around the magnetic background.
Thus, we substitute the plane wave superpositions for the linearized electromagnetic field, and discuss the wave propagation properties
in a material medium in the presence of the electric and magnetic conductivities.
From the wave equation, we calculate the refractive index solutions in terms of the magnetic background, of the EH parameters,
and of the electric/magnetic conductivities for the material medium. In this first case, the refractive index has a real and imaginary parts,
that are interpreted as the dispersion and the absorption of the wave, respectively. In the second case, the magnetic current density
for an isotropic magnetic conductivity is investigated in which there is no wave absorption. One fact is important : in all these
solutions, the refractive index depends on the $\theta$-angle that the magnetic background ${\bf B}$ does with the wave propagation direction
$\hat{{\bf k}}$, {\it i. e.}, $\cos\theta=\hat{{\bf B}}\cdot\hat{{\bf k}}$. Thereby, the solutions have different situations when
${\bf B}$ is parallel to ${\bf k}$, and when ${\bf B}$ perpendicular to ${\bf k}$. The nature of wave equations opens
the discussion of the birefringence phenomenon, that depends on the EH parameters, and on the magnetic background.
When the EH parameter is removed, the birefringence is null, and as well, all the results of the Maxwell ED are recovered.
In the birefringence analysis, we examine the refractive index solutions when the magnetic background is parallel $(n_{\parallel})$
and perpendicular $(n_{\perp})$ to the wave polarization direction. The birefringence emerges from the difference $\delta n=n_{\parallel}-n_{\perp}$, that provides a real and imaginary parts for a Ohmic conductivity $\sigma \neq 0$. For a weak magnetic background, we plot the real and imaginary parts of $\delta n$ as functions of the $\omega$-frequency. For high frequencies, the imaginary part goes to zero, whereas the real part goes to $\Re[\delta n]=3.27 \times 10^{-22}$, that is consistent known result from the PVLAS (polarization vacuum with laser) experiment for the vacuum birefringence $\Delta n/B^2=(19\pm 27) \times 10^{-24}\,\mbox{T}^{-2}$, when the magnetic background is $|{\bf B}|=2.5$ T. The solution of null birefringence for the real part $\Re[\delta n]=0$ is valid when $\omega=4\sigma/\sqrt{33}$. For end, this paper opens the discussion of applications of these refractive index solutions to others non-linear electrodynamics, as the ModMax ED. Other perspective is to investigate the effects of the linearization in optics classical law, where the medium is affected by the external magnetic field. These are discussions for a forthcoming project.

%

%
%\section{Acknowledgments}

%L.P.R. Ospedal is grateful to FAPERJ {\color{red} (grant number E-26/203.997/2022)} for his post-doctoral fellowship.
%P. Gaete was partially supported by ANID PIA/APOYO AFB220004.

%ANID PIA / APOYO AFB180002.

%\hspace{0.5cm}

%{\bf Data Availability Statement: No Data associated in the manuscript.}

\end{document}